\pdfoutput=1

\documentclass[11pt,twoside,a4paper,cmspaper,final,collab]{cms-tdr}
\def\svnVersion{433989P}\def\cmsCernNoTag{CERN-EP-2017-159}\def\cmsCernDate{\today}\def\cmsMessage{Published in the Journal of High Energy Physics as \href{http://dx.doi.org/10.1007/JHEP11(2017)010}{\doi{10.1007/JHEP11(2017)010.}}}

\begin{document}\cmsNoteHeader{HIG-15-009}

\hyphenation{had-ron-i-za-tion}
\hyphenation{cal-or-i-me-ter}
\hyphenation{de-vices}
\RCS$Revision: 370047 $
\RCS$HeadURL: svn+ssh://svn.cern.ch/reps/tdr2/notes/HIG-15-009/trunk/HIG-15-009.tex $
\RCS$Id: HIG-15-009.tex 370047 2016-10-06 13:05:37Z anikiten $

\cmsNoteHeader{HIG-15-009}
\title{Search for a light pseudoscalar Higgs boson produced in association with bottom quarks in pp collisions at
       $\sqrt{s}=8\TeV$}

\date{\today}

\abstract{
A search for a light pseudoscalar Higgs boson (\PXXA) produced in association with bottom quarks and decaying into a muon pair is reported. The search uses 19.7\fbinv of proton-proton collisions at a center-of-mass energy of 8\TeV, collected by the CMS experiment. No signal is observed in the dimuon mass range from 25 to 60\GeV. Upper limits on the cross section times branching fraction,
$\sigma(\Pp\Pp\to \bbbar\PXXA)\, \mathcal{B}(\PXXA\to\PGm\PGm)$, are set. }

\hypersetup{%
pdfauthor={CMS Collaboration},%
pdftitle={Search for a light pseudoscalar Higgs boson produced in association with bottom quarks in pp collisions at sqrt(s)=8 TeV},%
pdfsubject={CMS},%
pdfkeywords={CMS, physics, Higgs}}

\maketitle
\section{Introduction}
\label{sec:pas_introduction}

The CMS and ATLAS experiments at the CERN LHC have made clear observations of a particle
compatible with the standard
model (SM) Higgs boson~\cite{atlasHiggs, cmsHiggs, long, Aad:2015zhl}.
While the observation serves as a powerful affirmation
of the SM mechanism for generating particle masses, the discovery does not rule out the existence of a more
complex theory with an extended Higgs sector. The discovery of additional scalar
or pseudoscalar bosons may be evidence of such a scenario.

A number of well-motivated models extending the SM predict the existence of new Higgs bosons.
One well studied scenario is the Two Higgs Doublet Model (2HDM)~\cite{Gunion:1989we,Branco:2011iw}
predicting the existence of additional elementary Higgs particles. The 2HDM includes  a CP-odd Higgs boson, A,
that could be light and produced in association with bottom quarks at the LHC.
In some schemes, with a negatively-signed Yukawa coupling to the down-type fermions, the associated
$\bbbar\PXXA$ production cross section times branching fraction into muons can be very large for an
A boson with a mass below about 60\GeV, i.e. less than half of the mass of the observed SM-like Higgs
boson~\cite{Bernon:2014nxa}.  It can vary from 1000 to 35\unit{fb} for A boson masses in the range from 25 to
60\GeV at a center-of-mass energy of 8\TeV in proton-proton collisions.

We present a search for muon pairs, in the mass range from 25 to 60\GeV, produced in association
with bottom quarks in the context of the 2HDM. The search uses 19.7\fbinv of data collected by the
CMS experiment in proton-proton collisions at a center-of-mass energy of 8\TeV. The analysis of
the $\PGm\PGm$ final state complements the CMS searches for a low-mass A boson decaying into $\tau$ lepton
pairs by the CMS Collaboration ~\cite{Khachatryan:2015baw,Khachatryan:2017mnf}. Despite the significantly
lower branching fraction,
the $\PGm\PGm$ final state profits from much better dilepton mass resolution as well as higher trigger and lepton
identification efficiencies. This is the first time associated production with b quarks in the dimuon decay
channel has been searched for in the low dimuon mass range.

In the following, Sections 2 and 3 describe the CMS detector, the data, and simulated samples.
The event reconstruction and selections are presented in Section 4. In Section 5, the result of the search for
the $\bbbar\PXXA$ process is presented. The paper is summarized in Section 6.
\section{The CMS detector\label{sec:cms}}

\label{sec:pas_cms}
The central feature of the CMS apparatus~\cite{ref:cms} is a superconducting solenoid, 13\unit{m} in
length and 6\unit{m} in diameter, which provides an axial magnetic field of 3.8\unit{T}.
Within the field volume there are several particle detection systems. Charged-particle
trajectories are measured by silicon pixel and strip trackers, covering $0\le \phi \le
2\pi$ in azimuth and $\abs{\eta} < 2.5$ in pseudorapidity. A lead tungstate
crystal electromagnetic calorimeter (ECAL) surrounds the tracking volume. It is comprised
of a barrel region $\abs{\eta} < 1.48$ and two endcaps that extend up to $\abs{\eta} = 3$.
A lead and silicon-strip preshower detector
is located in front of the ECAL endcaps. A brass and scintillator hadron
calorimeter surrounds the ECAL and also covers the region $\abs{\eta} < 3$.  Steel
forward calorimeters with quartz fibers, read out by photomultipliers, extend the
calorimetric coverage up to $\abs{\eta} = 5$. Muons are identified and measured in
gas-ionization detectors embedded in the steel flux-return yoke outside the solenoid. The
detector is nearly hermetic, allowing energy balance measurements in the plane transverse
to the beam direction.
\section{Data and simulated samples}

A two-tier trigger system selects collision events of interest. For this analysis,
events are first selected by requiring a single muon ($\mu$) or two muons
within the detector acceptance and passing loose identification and kinematic requirements.
For the single-muon trigger, the muon must have $\abs{\eta} < 2.1$. The muon is required to be
isolated and must have transverse momentum ($\pt$) greater than 24\GeV. The  dimuon  trigger  requires 
two muons with \pt greater than 17\GeV for the leading muon and 8\GeV for the sub-leading muon.
The dielectron trigger  requires two electrons with \pt greater than 17\GeV for the leading electron 
and 8\GeV for the sub-leading electron and the electrons are required to be isolated.

The analysis uses opposite-sign dimuon events with additional jets, with at least one of them being
identified as originating from b quark fragmentation (``b jet").
The invariant mass of the opposite-sign lepton pair ($m_{\PGm\PGm}$) is required to be greater than
12\GeV to reject low-mass resonances and remove poorly-modeled backgrounds.

The Monte Carlo (MC) simulation is used to optimize the event selection to give the best sensitivity
for signal discovery. Predictions for the yields of background processes, which are also based on simulation,
are validated using data in appropriate control regions. The $\Pp\Pp \to \bbbar\PXXA$ signal events are
simulated with \PYTHIA (v6.4.26)~\cite{Sjostrand:2006za}. The signal samples have been generated for \PXXA masses
in intervals of 10\GeV. The natural width of the \PXXA boson has been set to a value of less than 50\MeV, which is
much smaller than the dimuon mass resolution in the mass range from 25 to 60\GeV
($\sigma_{\PGm\PGm}=0.45\GeV$ for $ m_{\PXXA}=30\GeV$).

The following background processes have been considered: Drell--Yan (DY), W+jets, \ttbar, single top
quark, and diboson production.

The DY events are simulated at leading order (LO) using the \MADGRAPH (v5.1.3.30)
generator~\cite{Alwall:2011uj,Alwall:2014hca} interfaced to \PYTHIA (v6.4.26)
for parton showering and hadronization. The CTEQ6L1~\cite{Pumplin:2002vw} parton distribution
functions (PDFs) are used to generate the events. The \MADGRAPH generation includes up to four partons in
the matrix element calculations and a procedure to properly merge jets from the matrix element calculations
and parton shower~\cite{Mangano:2006rw,Alwall:2007fs}.
The DY sample is normalized to the next-to-next-to-leading order (NNLO) cross section
computed with \FEWZ (v3.1)~\cite{Melnikov:2006kv}. The W+jets sample is simulated and normalized
in a similar way to the DY+jets sample. The $\ell\ell$ ($\ell \nu$) production in association with
b and c quarks is included in the DY+jets (W+jets) samples.

Top quark pair events are generated with \MADGRAPH, including up to three extra partons,
and the simulated sample is normalized to the NNLO+NNLL (next-to-next-to-leading-logarithmic)
inclusive cross section~\cite{Czakon:2013goa}. Single top quark processes (t, s, and Wt channels)
are modeled at next-to-leading order (NLO) with \POWHEG~\cite{Alioli:2010xd,Nason:2004rx,Frixione:2007vw}
and normalized to the approximate NNLO cross sections ~\cite{Kidonakis:2012db,Kidonakis:2013zqa}.
Diboson production processes (VV) $\PW\PW$, $\PW\Z/\gamma^{\ast}$ and
$\Z/\gamma^{\ast}\Z/\gamma^{\ast}$ are generated with \MADGRAPH
and normalized, respectively, to their NLO cross sections~\cite{Gehrmann:2014fva,Campbell:2010ff}.
The $\PW\Z/\gamma^{\ast}$ and $\Z/\gamma^{\ast}\Z/\gamma^{\ast}$ processes
are generated with $m_{\gamma^{\ast}}>10\GeV$.

The \PYTHIA parameters for the underlying event are set according to the Z2*
tune~\cite{Chatrchyan:2013gfi,Khachatryan:2015pea}, an update of
the Z1 tune described in Refs.~\cite{Field:2011iq, Field:2009zz}.

A detector simulation based on \GEANTfour (v.9.4p03)~\cite{Allison:2006ve,Agostinelli:2002hh}
is applied to all the generated signal and background samples.
The presence of multiple pp interactions (pileup) in the same or adjacent bunch crossings
is incorporated by simulating additional interactions (both in-time and out-of-time with the
collision) with a multiplicity distribution that matches the one observed in data.
The average number of pileup
events is estimated as 21 interactions per bunch crossing.
The observed jet energy resolution and jet energy scale measurements~\cite{ref:jetscale}, b tagging
efficiency and b tagging discriminator distributions~\cite{Chatrchyan:2012jua} in data are used to correct
the simulated events.

\section{Event reconstruction and selection}
\label{sec:pas_preselection}

Events are required to have at least one vertex, with the
reconstructed longitudinal position ($z$) within 24\unit{cm} of the
geometric center of the detector and the transverse position
within 2\unit{cm} of the beam interaction region.  In the case of
multiple reconstructed vertices associated with additional pp
interactions, the one with the highest scalar sum of the
${\pt^2}$ of its associated tracks is chosen as the primary vertex.
Muons are required to originate from the same primary vertex by
requiring the longitudinal and transverse impact parameters with respect
to the primary vertex to be less than 0.5\unit{cm} and 0.2\unit{cm}, respectively.

Muon candidates are reconstructed with a global trajectory fit using
hits in the tracker and the muon system~\cite{muon}. The efficiency for muons
to pass both the identification and isolation requirements is measured to be
more than 95\% for the kinematic region studied in this analysis~\cite{long}.

The electron selection criteria are optimized using a multivariate
approach, and have a combined identification and isolation efficiency of
approximately 60\% at low \pt ($\approx$10\GeV) and 90\% at high
\pt ($>$50\GeV) for electrons from $\PW$ or $\Z$ boson
decays~\cite{Khachatryan:2015hwa}. The training of the
multivariate electron reconstruction is performed using simulated
events, while the performance is validated using data.

In the interest of distinguishing between prompt and nonprompt lepton
candidates (predominantly arising from decays of b hadrons), a relative isolation
is defined for each lepton  candidate.
The muon isolation variable is computed as the sum of the transverse
momenta of the charged particles inside a
cone of radius  $\DR \equiv \sqrt{\smash[b]{(\Delta\eta)^2 + (\Delta
\phi)^2}} = 0.3$   around the lepton direction divided by the muon \pt.
The isolation is required to be smaller than 0.1. If the signal muons
are within a distance of $\DR < 0.3$ of each other
the \pt of one lepton is subtracted from the isolation value of the other lepton.

The anti-\kt clustering algorithm~\cite{Cacciari:2008gp,Cacciari:2011ma}
with a distance parameter of 0.5 is used for jet reconstruction.

Two types of jets are used in the analysis, ``jet-plus-track" (JPT) ~\cite{CMS-PAS-JME-14-005} and
particle-flow (PF)~\cite{Sirunyan:2017ulk}
jets.
The two algorithms combine the information from all subdetectors.
However, the PF algorithm separates deposits in the calorimeter into clusters and matches clusters
with reconstructed tracks (PF candidates), while the JPT algorithm does not require cluster separation as it
starts from the reconstructed calorimeter jets (calojet) and track matching is done with respect to
the calojet area at the surface of the calorimeter. Also
the algorithms have a different approach to the calibration of the constituents: cluster calibration from
simulation in the case of the PF algorithm; and corrections due to calorimeter thresholds and tracking inefficiency
in the case of the JPT algorithm. These jet reconstruction algorithms are affected differently by detector effects,
like tracker misalignment or the calibration of the calorimeter, and thus serve as an important
cross-check of each other. The two types of jet, JPT and PF, provide similar jet energy resolution.
The final results are presented with PF jets.

Jets with a significant fraction of energy coming from
pileup interactions or not associated with the primary vertex are
rejected~\cite{CMS-PAS-JME-13-005,CMS-PAS-JME-14-005}. The remaining pileup energy in
jets is subtracted using a technique that relies on information
about the jet area~\cite{pile1,pile2,Cacciari:2011ma}. Jets are required to satisfy identification criteria
that remove jets originating from noisy channels in the calorimeters~\cite{ref:met,ref:noise}.
Calibrated jets~\cite{ref:jetscale,Khachatryan:2016kdb} are required to have $\pt >
20$\GeV, $\abs{\eta} < 4.7$ and to be separated by at least 0.5 in $\DR$ from muons passing the
selection requirements described above.

The missing transverse momentum ($\ptmiss$) is defined as the modulus of the vector sum of the transverse
momenta of all reconstructed PF candidates in the event.
The calibrations applied to jets are propagated to the $\ptmiss$.

Jets originating from the hadronization of $\PQb$ (or $\PQc$) quarks
are identified by a multivariate analysis ~\cite{Hocker:2007ht}  based on a
secondary vertex b tagging algorithm described in Refs.~\cite{Chatrchyan:2012jua,CMS-PAS-BTV-13-001}.
The algorithm combines information on track impact parameter significance, secondary vertices, and
jet kinematics. Several working points are available; these are designed to yield successively higher
purity at the cost of reduced efficiency for identifying b jets.
The working point used in this analysis corresponds to a b tagging identification efficiency of 45\%
and a misidentification rate of 0.05\% for light-flavor (u, d, s, g) jets with $\pt > 30\GeV$. The
misidentification rate for c jets is $\simeq$6\%.

Selections are optimized using simulated events for the signal model
$\Pp\Pp\to\bbbar\PXXA,\ \PXXA\to \PGm\PGm$ with
$m_{\PXXA}=30\GeV$ and simulated backgrounds. The threshold on the leading muon
$\pt^{\mu_{1}}$ is fixed to 25\GeV. This choice is determined by the single-muon trigger
\pt threshold of 24\GeV. The rapidity requirements on the muons and b jets are driven by the trigger and
particle identification conditions. The optimization procedure selects the thresholds on the
subleading \pt muon ($\pt^{\mu_{2}}$), the leading b jet \pt ($\pt^{\text{\PQb jet}}$),
and $\ptmiss$, where an approximate estimate of signal significance, defined as
$\Z = 2 (\sqrt{S+B}-\sqrt{B})$~\cite{Bityukov:1999sa} reaches its maximal value.
The $\ptmiss$ requirement is used to suppress the $\ttbar$ background.
The presence of a second jet (b tagged or not) is not required since in the $\bbbar\PXXA$
production process the second b jet is often at high $\eta$ or has too low \pt.

Optimization in this way gives rise to:
\begin{itemize}
\item $\pt^{\mu_1}>25$\GeV, $\abs{\eta_{\mu_1}}<2.1$;
\item $\pt^{\mu_2}>5$\GeV, $\abs{\eta_{\mu_2}}<2.4$;
\item $\pt^{\text{b jet}}>20$\GeV and $\abs{\eta}<2.4$;
\item $\ptmiss<40$\GeV.
\end{itemize}

The optimization of the selections was repeated for an A boson mass of 60\GeV and similar values
were obtained. The signal selection efficiency is 0.013 for $m_{\PXXA}=30\GeV$.
\section{Results and systematic uncertainties \label{sec:results}}

\begin{figure}[hbtp]
  \centering
{\includegraphics[width=0.48\textwidth]{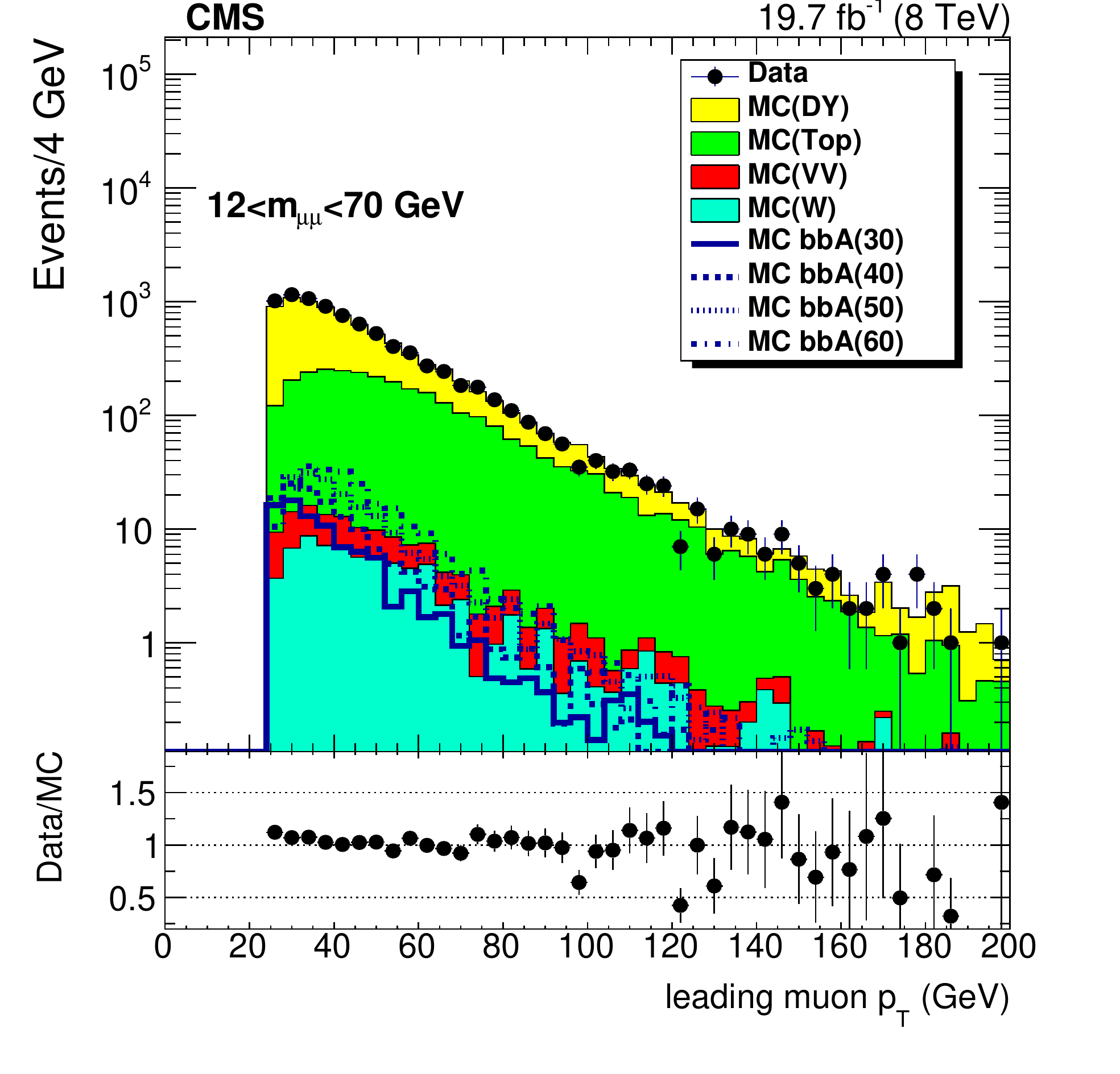}}
    \hfill
{\includegraphics[width=0.48\textwidth]{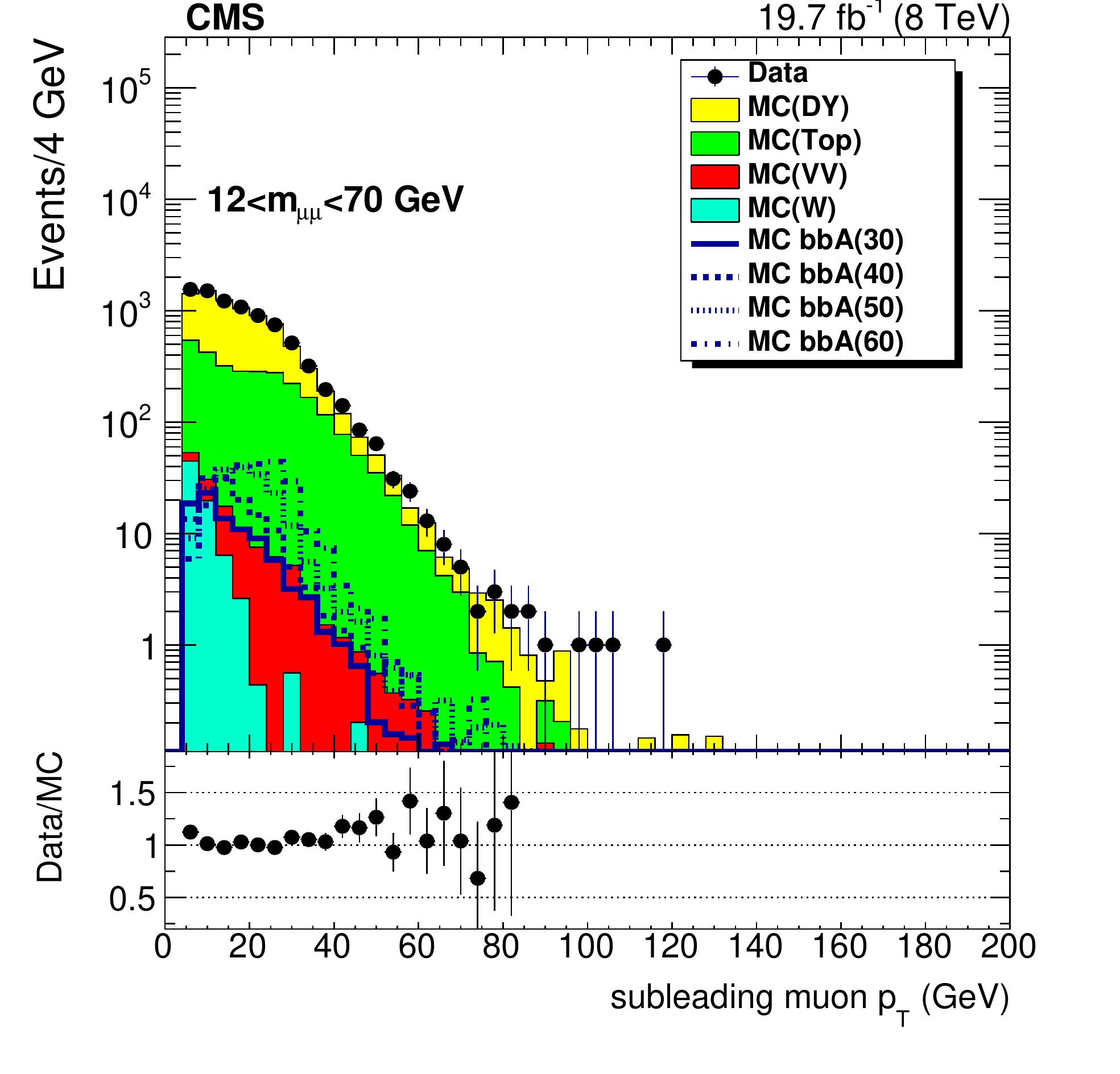}}
     \caption{
       The transverse momentum of the leading (left) and the subleading (right)
       muon for data (dots) and simulation (histograms). The histograms for
       simulated backgrounds are stacked. The histogram labeled as MC(Top) shows the sum of the single top
       and top quark pair production processes. The expected signal is shown assuming a signal cross
       section times branching fraction of 350\unit{fb}. The background is normalized to the number of
       events expected from simulation. In the lower panel, the ratio of the number of events
       in data and background simulation is shown.}
    \label{fig:ptmuons}
\end{figure}

\begin{figure}[hbtp]
  \centering
{\includegraphics[width=0.48\textwidth]{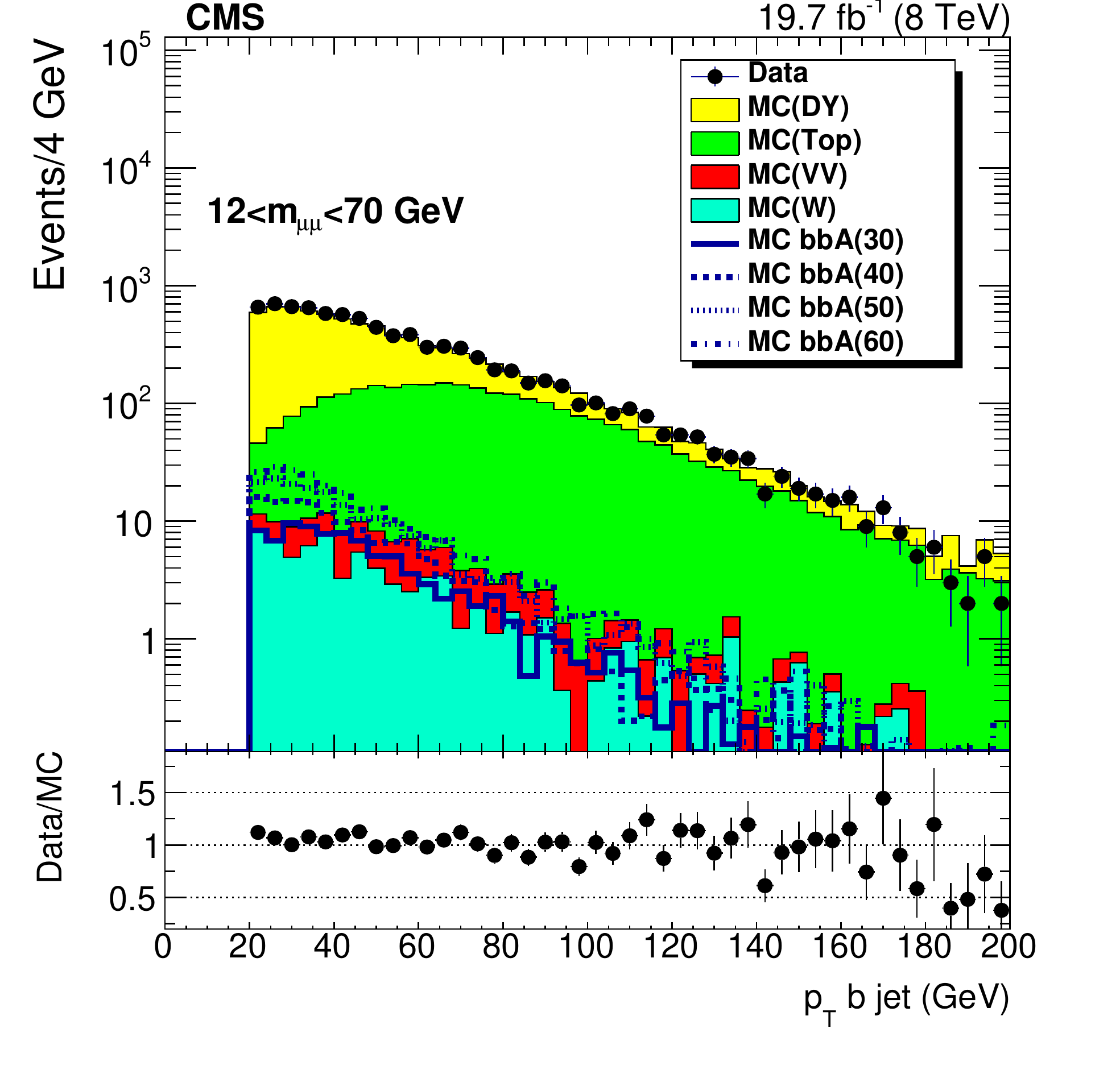}}
    \hfill
{\includegraphics[width=0.48\textwidth]{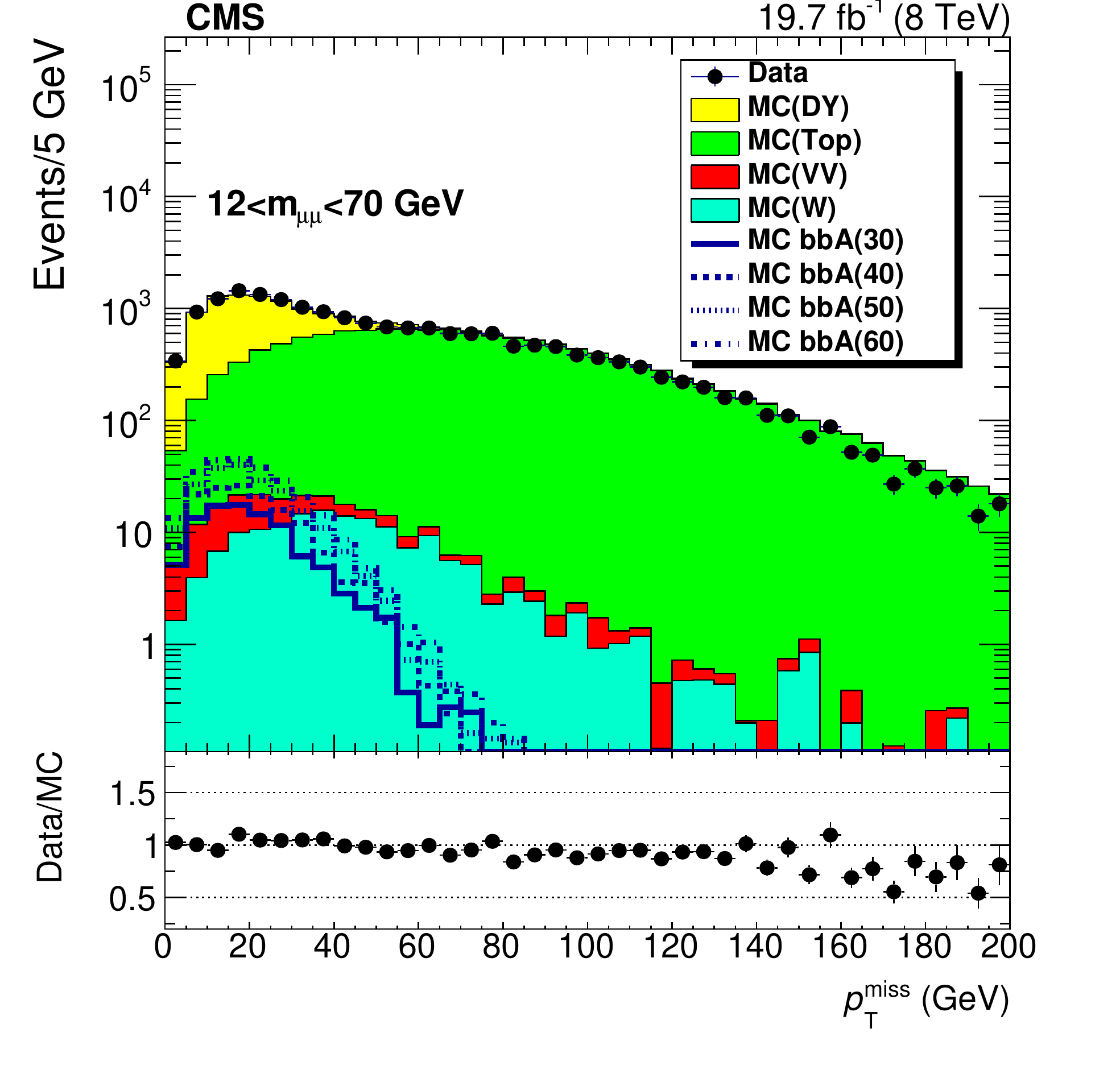}}
     \caption{The transverse momentum of the leading \pt b jet (left) and
              the missing transverse momentum (right) for data (dots) and simulation (histograms).
              The histograms for simulated backgrounds are stacked. The histogram labeled as MC(Top)
              shows the sum of the single top and top quark pair production processes.
              The expected signal is shown assuming a signal
              cross section times branching fraction of 350\unit{fb}. The background is normalized to the number
              of events expected from simulation. In the lower panel, the ratio of the number of events
              in data and background simulation is shown.}
    \label{fig:ptbmet}
\end{figure}

Figure~\ref{fig:ptmuons} shows the transverse momentum of the leading (left) and subleading (right)
muons, and Fig.~\ref{fig:ptbmet} shows the \pt of the leading b jet (left)
and the $\ptmiss$ (right) for the events passing the selection requirements for the other variables.
The histogram labeled as MC(Top) shows the sum of the single top and top quark pair production processes.

The upper limits on the signal contribution in the mass range of $25 < m_{\PGm\PGm}<60\GeV$
have been extracted using a fit of the binned background and signal templates to the dimuon mass distribution
in the region $12 < m_{\mu\mu}<70\GeV$. The mass intervals of [12--25] and [60--70]\GeV have
been used for a verification of the background expectation obtained from the simulation.

The signal and background templates are obtained from simulation rescaled and corrected using information
from data. To obtain signal mass shapes in 1\GeV steps we perform a linear interpolation between histograms
as a function of the mass using the algorithm described in Ref.~\cite{Read:1999kh} and implemented in the
\textsc{root} package ~\cite{Brun:1997pa}. The W+jets and VV background event yields are found to be
negligible. The predicted yields for the background processes take into account the following sources of
systematic uncertainty, where the range of variations of each source was determined in dedicated studies
with appropriate data control samples or, if relevant, from the theory uncertainty affecting the prediction:
\begin{itemize}
\item luminosity uncertainty, 2.6\%~\cite{LUM-13-001};
\item muon identification and isolation efficiency uncertainty, 3\%;
\item top quark background normalization uncertainty, 7\%;
\item jet energy scale uncertainty, 3\% for the DY+jets and 0.2\% for top quark backgrounds;
\item jet energy resolution uncertainty, 0.3\% for DY+jet and 0.1\% for top quark backgrounds;
\item pileup modeling uncertainty, 2\% for DY+jets and 1\% for top quark backgrounds;
\item $\ETmiss$ measurement uncertainty, 2\% for DY+jets and 4\% for top quark backgrounds.
\end{itemize}

While the above uncertainties only affect the expected total event yield, others
can have an impact both on the total event yield and on the shape of the predicted
$m_{\PGm\PGm}$ distribution. The following set of systematic uncertainties in the
$m_{\PGm\PGm}$ background shape and the normalization have been taken into account:
\begin{itemize}
\item b tagging scaling factor uncertainty~\cite{CMS-PAS-BTV-13-001};
\item renormalization and factorization scale, and PDF uncertainties;
\end{itemize}

The renormalization and factorization scale uncertainties are particularly important for the modeling of the shape of
the DY+jets background, which after the final selections is dominated by the $\PGm\PGm \bbbar$ process.
The scale uncertainty in the $\PGm\PGm \bbbar$ prediction, taken from simulation, has been
evaluated as a function of the dimuon mass using the $\Pp\Pp\to \PGm\PGm \bbbar$ process
implemented in \MGvATNLO v2.3.0 in a four-flavor production scheme with
massive b quarks~\cite{Frederix:2011qg}. The renormalization and factorization scales
($\mu_\mathrm{R}$ and $\mu_\mathrm{F}$) have been varied simultaneously by
factors of 0.5 and 2.0. The uncertainty on the differential cross-section as a function of the dimuon mass
varies from 20 to 15\% in the dimuon mass region of $12 < m_{\PGm\PGm}<70\GeV$.
It is applied to the total DY+jets background, motivated
by the fact the DY plus light-flavor jets background after the b tagging requirement is much smaller
than the $\PGm\PGm \bbbar$ background, and the scale uncertainty for the $\PGm\PGm \ccbar$ process
is expected to be similar to that for the $\PGm\PGm \bbbar$ process.

Uncertainties related to the imprecision in our knowledge of the PDFs are also evaluated.
The uncertainty in the gluon-gluon luminosity is considered, since in the four-flavor scheme 
the dominant LO production process is $\Pg\Pg \to \PGm\PGm \bbbar$. The $\qqbar \to \PGm\PGm \bbbar$
process represents only $\simeq$15\% of the total cross-section.
The gluon-gluon luminosity uncertainty is compared for three PDF sets, CT10NLO~\cite{Lai:2010vv},
MSTW2008NLO~\cite{Martin:2010db}, and NNPDF2.3NLO~\cite{Ball:2012cx}.
In the mass region of $\simeq$30--70\GeV the CT10NLO uncertainty envelope covers the uncertainties
of the other two PDF sets, therefore the CT10NLO gluon-gluon luminosity uncertainty is used in the analysis.
This PDF uncertainty is applied as a function of the dimuon mass to the total DY+jets background after
b tagging has been applied. It is motivated by the fact that the $\PGm\PGm \ccbar$ and $\PGm\PGm \bbbar$
processes are both initiated by gluon-gluon interactions in the four-flavor scheme at LO.

\begin{figure}[htp]
  \centering
{\includegraphics[width=0.48\textwidth]{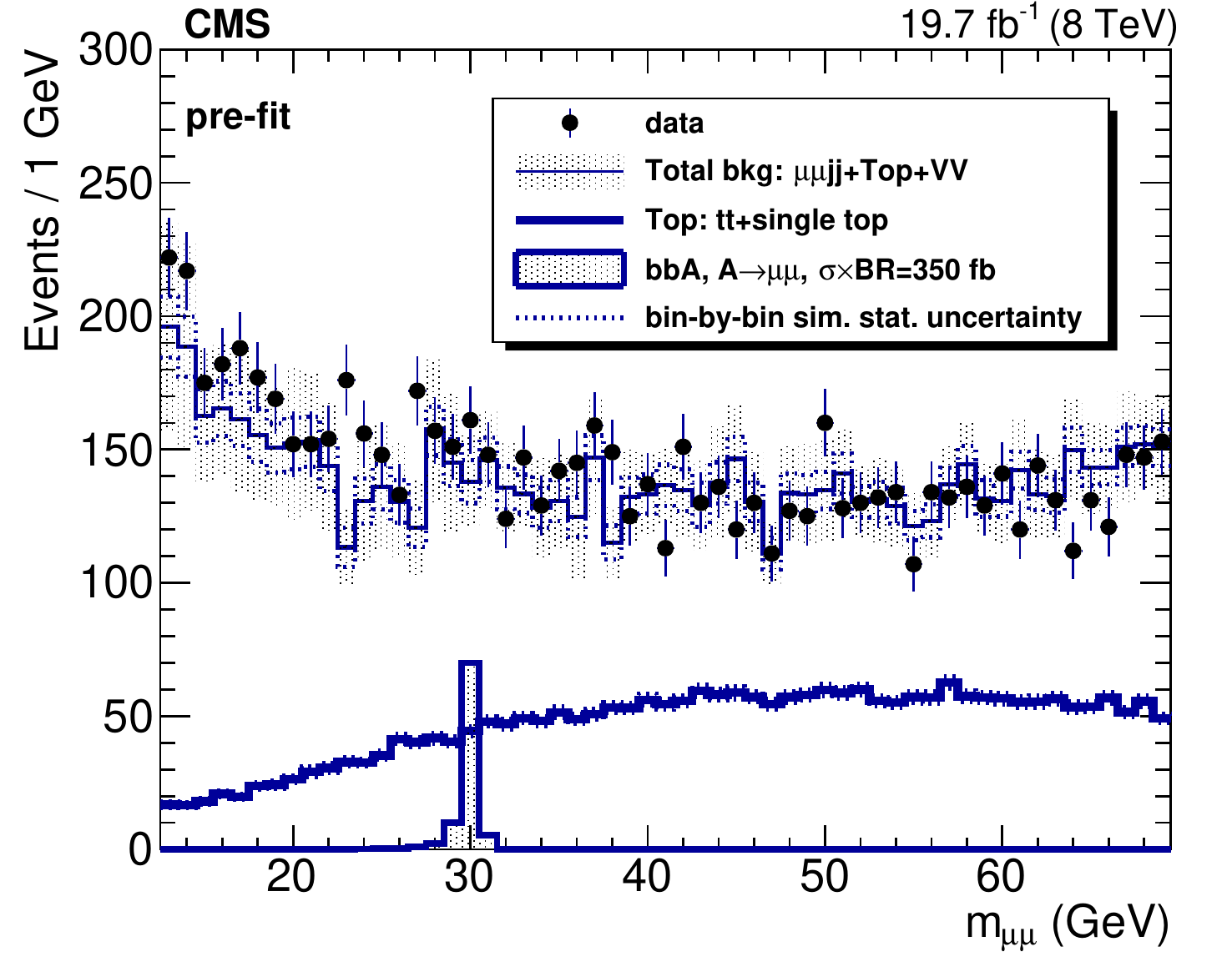}}
{\includegraphics[width=0.48\textwidth]{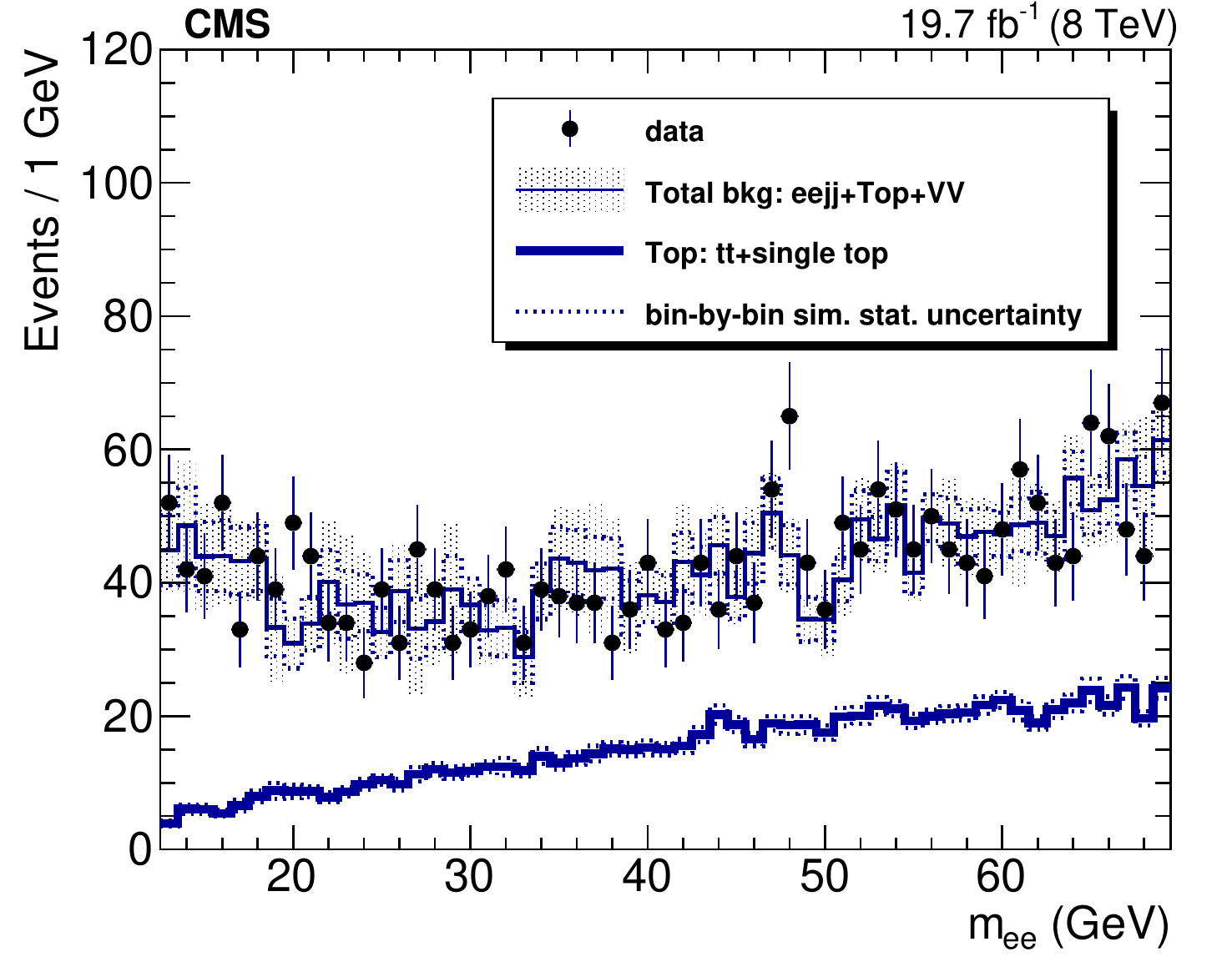}}
    \caption{Left: the dimuon mass distribution with the pre-fit expected background event yield and its uncertainty,
      and the expected signal for $m_{\PXXA}=30\GeV$ assuming a signal cross section
      times branching fraction of 350\unit{fb}. Right: cross-check with the $\Pep\Pem$ final state
      showing the dielectron mass spectrum with the expected background event yield and its uncertainty.}
    \label{fig:10_70}
\end{figure}

The dimuon mass spectrum, the expected background, and its uncertainty are presented in
Fig.~\ref{fig:10_70} (left) together with the expected signal for $m_{\PXXA}=30\GeV$ assuming a
signal cross section times branching fraction of 350\unit{fb}. The PF jet reconstruction algorithm
is used. There is a fair agreement between data and background.

To cross-check the background estimation technique that relies on the accuracy of
the simulation for the background, we consider data where instead of requiring the presence 
of two muons, we require two opposite-sign electrons. Since the dielectron final state would be 
suppressed by $({m_{\Pe}}/{m_{\PGm}})^2$ in the pseudoscalar Higgs decay, 
we do not consider it for our search. Kinematic thresholds and all other requirements are kept 
exactly the same as in the main analysis, except the threshold on the second leading \pt electron, 
which is taken higher than in the dimuon analysis ($\pt^{\Pe_{2}} > 10\GeV$) to reduce the QCD multijet 
background. The data are found to be in good agreement with the background model 
prediction, as illustrated in Fig.~\ref{fig:10_70} (right).

\begin{figure}[htp]
  \centering
{\includegraphics[width=0.48\textwidth]{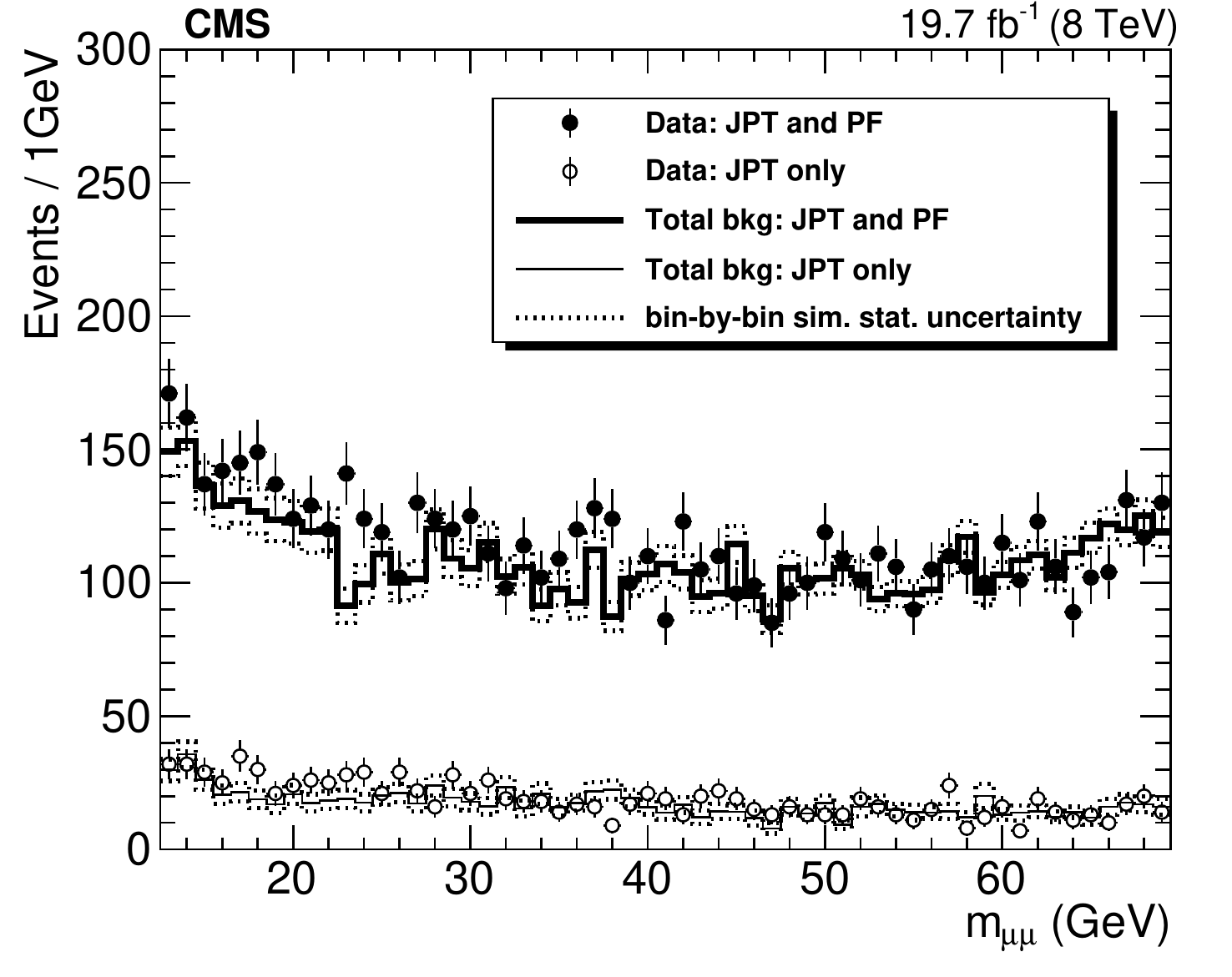}}
    \caption{The dimuon mass distribution with the pre-fit expected background event yield for events selected by both the main (PF) and alternative (JPT) jet reconstruction methods
             as well as events selected by only the JPT algorithm.}
    \label{fig:JPT}
\end{figure}

To cross-check the effect of event migration on the exclusion limits due to the choice of jet reconstruction
algorithm we use the JPT algorithm as an alternative and repeat the full analysis chain.
Figure~\ref{fig:JPT} shows the dimuon mass spectrum for events selected by both the main (PF)
and alternative (JPT) jet reconstruction methods as well as for events selected by only the JPT
algorithm. The two reconstruction methods select slightly different events. The fraction of migrated events
due to the choice of jet algorithm is of the order of 20\% and almost independent of the dimuon mass.
The expected and observed upper limits obtained using PF or JPT jets are very similar. The final results presented
use PF jets.

The experimental uncertainties considered above for the background are also taken into
account for the signal. In addition, the uncertainty in the signal acceptance due to the
signal modeling is evaluated using a \MGvATNLO{}+\PYTHIA8 particle level simulation of the
$\Pp\Pp\to \bbbar\PXXA$, $\PXXA\to\PGm\PGm$
process in the four-flavor production scheme~\cite{Wiesemann:2014ioa,Dittmaier:2003ej,Dawson:2003kb}.
It includes the uncertainty in the shower scale parameter, $\alpha$ (an upper scale
allowed for the shower in \PYTHIA)~\cite{Alwall:2014hca} , the renormalization and factorization scales,
and the PDFs. In the evaluation of the uncertainties we follow the recommendations of
LHC Higgs Cross Section Working Group~\cite{deFlorian:2016spz}.
The renormalization and factorization scales are varied with
two constraints such that $0.5\mu_{0}\leq\mu_\mathrm{R},\mu_\mathrm{F}\leq 2\mu_{0}$
varies while requiring $0.5 \leq \mu_\mathrm{R} / \mu_\mathrm{F} \leq 2$ ($\mu_{0}$ is the recommended central scale
value~\cite{deFlorian:2016spz}); the shower scale parameter is varied as $\alpha \in [1/(4\sqrt{2}),\sqrt{2}/4]$.
The PDF uncertainty is obtained with the PDF4LHC15\_nlo\_nf4\_30 set
~\cite{Butterworth:2015oua,Ball:2014uwa,Dulat:2015mca,Harland-Lang:2014zoa}.
The total theoretical uncertainty on the signal acceptance, taken as the linear sum of the shower scale,
the renormalization and factorization scale, and the PDF uncertainties, varies from $-16$\% to $+18$\%.
The difference in the signal acceptance between the {\PYTHIA6} and \MGvATNLO{}+{\PYTHIA8}
particle level simulation is less than the total theoretical uncertainty discussed above.

\begin{figure}[htp]
  \centering
{\includegraphics[width=0.48\textwidth]{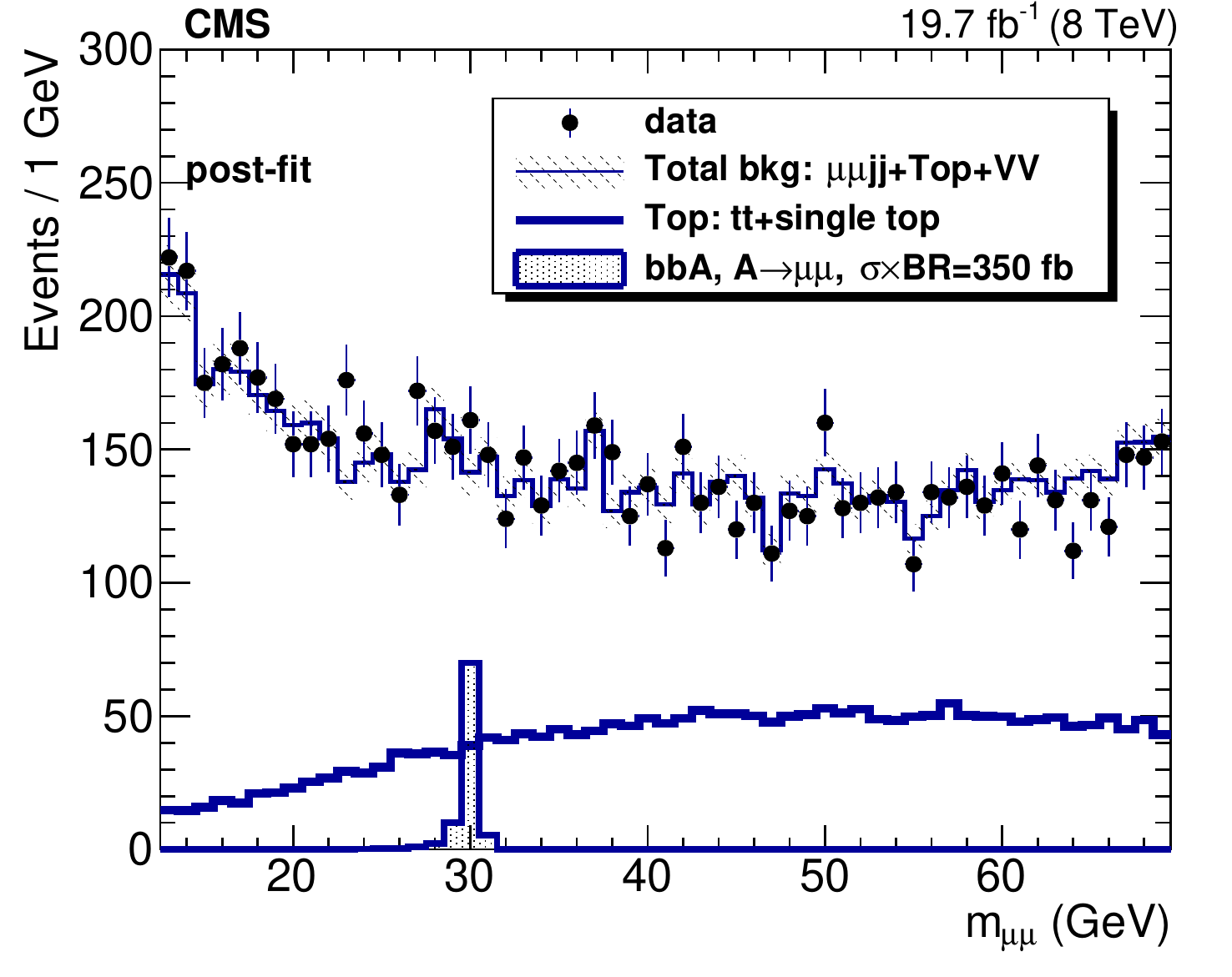}}
    \hfill
{\includegraphics[width=0.48\textwidth]{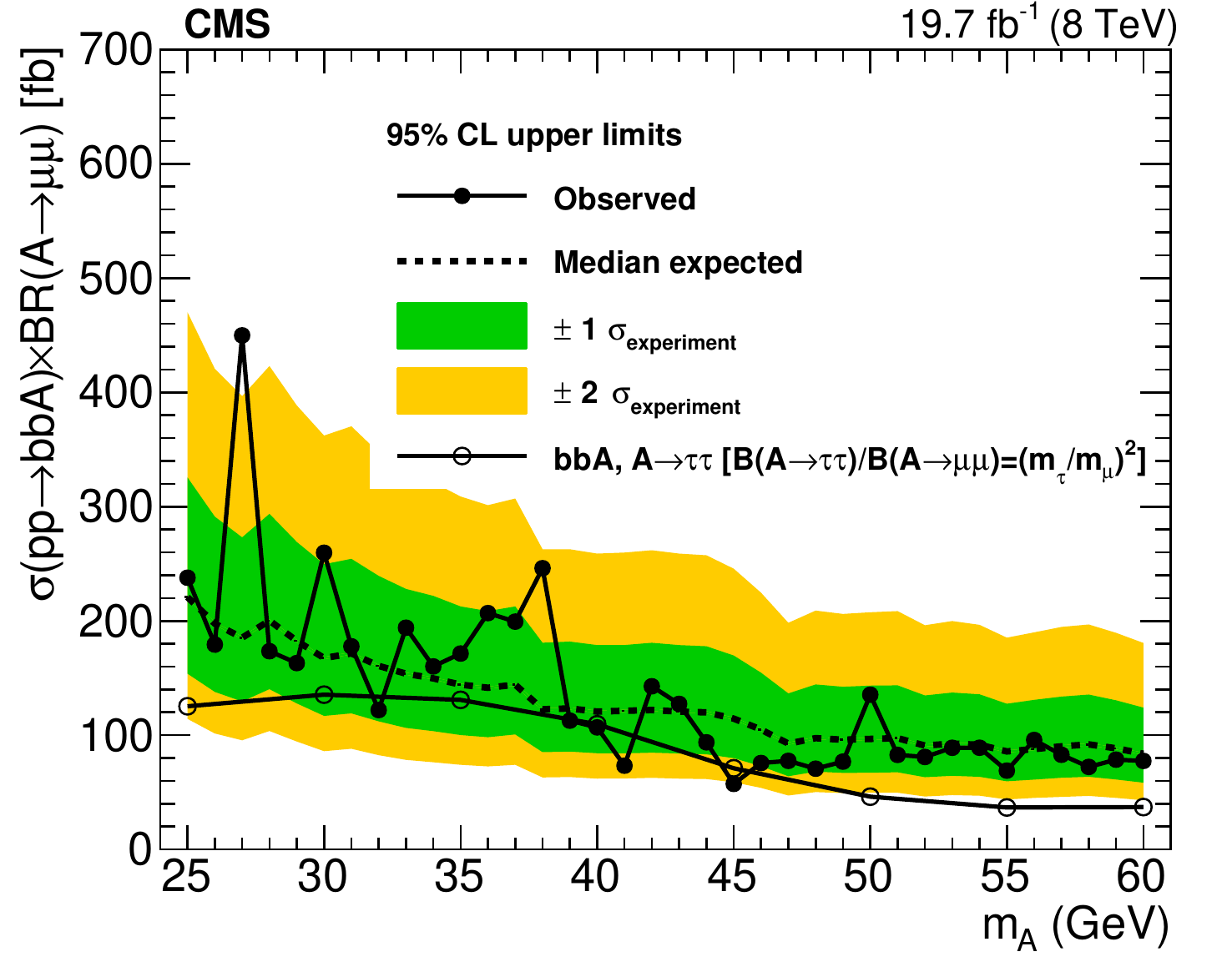}}
    \caption{Left: the dimuon mass distribution with the post-fit background event yield
              and its uncertainty given by the fit, and the expected signal for $m_{\PXXA}=30\GeV$ assuming a
              signal cross section times branching fraction of 350\unit{fb}.
             Right: expected and observed upper limit at 95\% CL on
              $\sigma(\Pp\Pp \to \bbbar\PXXA)\, \mathcal{B} \PXXA\to\PGm\PGm)$
              as a function of $m_{\PXXA}$. The open circles show the limits obtained in the CMS
              analysis of the $\PXXA\to\tau\tau$ final state~\cite{Khachatryan:2015baw}
              when translated into limits for the $\PXXA\to\PGm\PGm$ final state
              using Eq.(~\ref{eq:br}).}
    \label{fig:limits}

\end{figure}

The dimuon mass distribution in Fig.~\ref{fig:10_70} (left) is used in the evaluation
of the limits on the cross section times branching fraction,
$\sigma(\Pp\Pp \to \bbbar\PXXA)\, \mathcal{B}(\PXXA\to\PGm\PGm)$.
The limits have been extracted using a test statistic based on the profile likelihood ratio
and asymptotic formulae, along with the constraint of a positive signal, as proposed in
Refs.~\cite{Cowan:2010js,CMS-NOTE-2011-005}.
Systematic uncertainties are included in the form of nuisance parameters.
The dimuon mass spectrum and the post-fit background event yield and
its uncertainty given by the fit is presented in Fig.~\ref{fig:limits} (left), along with the expected
signal for $m_{\PXXA}=30\GeV$ assuming a signal cross section times branching fraction of 350\unit{fb}.

We scan over \PXXA masses from 25 to 60\GeV.
The expected and observed upper limits at 95\% confidence level (CL)  on
$\sigma(\Pp\Pp \to \bbbar\PXXA)\, \mathcal{B} (\PXXA\to\PGm\PGm)$
are shown in Fig.~\ref{fig:limits} (right). Due to the good agreement between
the data and the expected background before and after the fit the limits for all mass points in the mass
interval of $25 < m_{\PGm\PGm} < 60\GeV$ are within two standard deviations of the expected limit.

In the 2HDM, the $\mathcal{B} (\PXXA\to \PGm\PGm)$ and
$\mathcal{B} (\PXXA\to \tau\tau)$ are related through the ratio of the
Yukawa couplings, and thus are proportional to the square of the lepton masses, i.e.
\begin{equation}
\label{eq:br}
\frac{\mathcal{B}(\PXXA\to\tau\tau)}{\mathcal{B}(\PXXA\to\PGm\PGm)}=\Bigl(\frac{m_{\tau}}{m_{\mu}}\Bigr)^{2}.
\end{equation}

The open circles in Fig.~\ref{fig:limits} show the upper limits on
$\sigma(\Pp\Pp \to \bbbar\PXXA)\, \mathcal{B}(\PXXA\to\PGm\PGm)$ obtained
from the limits on $\sigma(\Pp\Pp \to \bbbar\PXXA)\, \mathcal{B}(\PXXA\to\tau\tau)$
evaluated in the CMS analysis of the $\PXXA\to\tau\tau$ final state~\cite{Khachatryan:2015baw}.
One can see that upper limits evaluated from the direct search for the $\PXXA\to\PGm\PGm$ decay
in $\bbbar\PXXA$ associated production are comparable with those from the $\PXXA\to\tau\tau$
search using the same production process.
\section{Summary}
\label{sec:pas_conclusions}

A light pseudoscalar Higgs boson, produced in association with a pair
of b jets and decaying into two muons, has been searched for
in pp collisions at $\sqrt{s}=8\TeV$ with an integrated luminosity of
19.7\fbinv. This is the first time associated production with b quarks in the dimuon decay
channel has been looked for in the low dimuon mass range.
No signal has been observed in the dimuon mass range from
25 to 60\GeV. Upper limits on the cross section times branching fraction,
$\sigma(\Pp\Pp \to \bbbar\PXXA)\, \mathcal{B}(\PXXA\to\PGm\PGm)$,
have been set. Despite the significantly lower branching fraction, the limits evaluated from the direct search
for the $\PXXA\to\PGm\PGm$ decay in $\bbbar\PXXA$ associated production are comparable with
those from the $\PXXA\to\tau\tau$ search using the same production process.

\begin{acknowledgments}
We congratulate our colleagues in the CERN accelerator departments for the excellent performance of the LHC and thank the technical and administrative staffs at CERN and at other CMS institutes for their contributions to the success of the CMS effort. In addition, we gratefully acknowledge the computing centers and personnel of the Worldwide LHC Computing Grid for delivering so effectively the computing infrastructure essential to our analyses. Finally, we acknowledge the enduring support for the construction and operation of the LHC and the CMS detector provided by the following funding agencies: BMWFW and FWF (Austria); FNRS and FWO (Belgium); CNPq, CAPES, FAPERJ, and FAPESP (Brazil); MES (Bulgaria); CERN; CAS, MoST, and NSFC (China); COLCIENCIAS (Colombia); MSES and CSF (Croatia); RPF (Cyprus); SENESCYT (Ecuador); MoER, ERC IUT, and ERDF (Estonia); Academy of Finland, MEC, and HIP (Finland); CEA and CNRS/IN2P3 (France); BMBF, DFG, and HGF (Germany); GSRT (Greece); OTKA and NIH (Hungary); DAE and DST (India); IPM (Iran); SFI (Ireland); INFN (Italy); MSIP and NRF (Republic of Korea); LAS (Lithuania); MOE and UM (Malaysia); BUAP, CINVESTAV, CONACYT, LNS, SEP, and UASLP-FAI (Mexico); MBIE (New Zealand); PAEC (Pakistan); MSHE and NSC (Poland); FCT (Portugal); JINR (Dubna); MON, RosAtom, RAS, RFBR and RAEP (Russia); MESTD (Serbia); SEIDI, CPAN, PCTI and FEDER (Spain); Swiss Funding Agencies (Switzerland); MST (Taipei); ThEPCenter, IPST, STAR, and NSTDA (Thailand); TUBITAK and TAEK (Turkey); NASU and SFFR (Ukraine); STFC (United Kingdom); DOE and NSF (USA).

{\tolerance=800
\hyphenation{Rachada-pisek} Individuals have received support from the Marie-Curie program and the European Research Council and Horizon 2020 Grant, contract No. 675440 (European Union); the Leventis Foundation; the A. P. Sloan Foundation; the Alexander von Humboldt Foundation; the Belgian Federal Science Policy Office; the Fonds pour la Formation \`a la Recherche dans l'Industrie et dans l'Agriculture (FRIA-Belgium); the Agentschap voor Innovatie door Wetenschap en Technologie (IWT-Belgium); the Ministry of Education, Youth and Sports (MEYS) of the Czech Republic; the Council of Science and Industrial Research, India; the HOMING PLUS program of the Foundation for Polish Science, cofinanced from European Union, Regional Development Fund, the Mobility Plus program of the Ministry of Science and Higher Education, the National Science Center (Poland), contracts Harmonia 2014/14/M/ST2/00428, Opus 2014/13/B/ST2/02543, 2014/15/B/ST2/03998, and 2015/19/B/ST2/02861, Sonata-bis 2012/07/E/ST2/01406; the National Priorities Research Program by Qatar National Research Fund; the Programa Clar\'in-COFUND del Principado de Asturias; the Thalis and Aristeia programs cofinanced by EU-ESF and the Greek NSRF; the Rachadapisek Sompot Fund for Postdoctoral Fellowship, Chulalongkorn University and the Chulalongkorn Academic into Its 2nd Century Project Advancement Project (Thailand); and the Welch Foundation, contract C-1845.
\par}
\end{acknowledgments}
\newpage
\bibliography{auto_generated}

\cleardoublepage \appendix\section{The CMS Collaboration \label{app:collab}}\begin{sloppypar}\hyphenpenalty=5000\widowpenalty=500\clubpenalty=5000\textbf{Yerevan Physics Institute,  Yerevan,  Armenia}\\*[0pt]
A.M.~Sirunyan, A.~Tumasyan
\vskip\cmsinstskip
\textbf{Institut f\"{u}r Hochenergiephysik,  Wien,  Austria}\\*[0pt]
W.~Adam, F.~Ambrogi, E.~Asilar, T.~Bergauer, J.~Brandstetter, E.~Brondolin, M.~Dragicevic, J.~Er\"{o}, M.~Flechl, M.~Friedl, R.~Fr\"{u}hwirth\cmsAuthorMark{1}, V.M.~Ghete, J.~Grossmann, J.~Hrubec, M.~Jeitler\cmsAuthorMark{1}, A.~K\"{o}nig, N.~Krammer, I.~Kr\"{a}tschmer, D.~Liko, T.~Madlener, I.~Mikulec, E.~Pree, D.~Rabady, N.~Rad, H.~Rohringer, J.~Schieck\cmsAuthorMark{1}, R.~Sch\"{o}fbeck, M.~Spanring, D.~Spitzbart, W.~Waltenberger, J.~Wittmann, C.-E.~Wulz\cmsAuthorMark{1}, M.~Zarucki
% ...
\vskip\cmsinstskip
% ...
1:~~Also at Vienna University of Technology, Vienna, Austria\\
% ...

\end{sloppypar}
\end{document}